# Site-resolved Bragg scattering

**Birjoo Vaishnav and David S. Weiss[*]**

*Physics Department, The Pennsylvania State University, University Park, PA 16802, USA*

[*]*Corresponding author: dsweiss@phys.psu.edu*

We numerically calculate the reliability with which one can optically determine the presence or absence of an individual scatterer in a randomly occupied 3D array of well-localized, coherently radiating scatterers. The reliability depends in a complicated way on the ratio of lattice spacing to wavelength and the numerical aperture of the imaging system. The behavior can be qualitatively understood by considering the dependence of Bragg scattering modes on lattice spacing. These results are of interest for atomic implementations of quantum information processing. © 2007 Optical Society of America

Neutral atoms trapped in a far-off-resonant optical lattice have been proposed as the basis for a quantum computer [1-3]. A quantum computation must start from a well known initial distribution of atoms among lattice sites. One approach is to directly observe the location of atoms in the lattice, and then use the information to either fill in the vacancies or to account for the vacancies in a customized computation [4,5]. Since it is easier to address individual atoms at lattice sites when the lattice constant $\ell$ is large, initialization and site addressability can be optimized together in this approach. However, a smaller $\ell$ makes it easier to mutually entangle atoms, and more of them can be fit into a smaller volume. Therefore, an important issue in the design of a 3D lattice-based quantum computer is the minimum size of $\ell$ consistent with the ability to individually resolve atoms.

Arrays of 250 single atoms in a 3D optical lattice have recently been imaged [6]. Compared to 1D and 2D single ion and atom arrays [7-10], more qubits can be trapped with more near neighbors in 3D. But 3D imaging presents the problem of background light from atoms outside of the image plane. The considerable prior study of light scattering from 3D arrays of particles has always been in momentum space, i.e., using Bragg scattering, where detailed information about vacancies is not available [11-13]. The demands of single site resolution require new theoretical studies of how best to maximize the signal to background. Although Bragg scattering concerns the interference of indistinguishable scatterers, it turns out to be relevant to interference affects in site-resolved images.

In this letter, we consider the problem of imaging individual light scatterers in a randomly occupied, finite, 3D lattice. We assume that the scatterers are localized to much better than a wavelength, which was not the case in Ref. [6], but might be attained for atoms with better laser cooling. We primarily consider site occupations of 50%, where a random half of the sites have exactly one scatterer [14,15]. The scatterers are coherently illuminated by a wavelength $\lambda$ traveling wave perpendicular to the imaging optical axis, linearly polarized parallel to the image plane. For a large set of atom distributions and assuming diffraction limited optics, we calculate the intensity in the image plane both with and without a particular target site being filled. The results determine the reliability with which we can identify a vacancy when the overall distribution is unknown.

The geometry for this calculation is shown in Fig.1. The optic axis is the $z$-axis. We define a new variable for the lattice spacing $\rho = \ell/\lambda$, and implicitly scale all other length variables by $\lambda$ to make them dimensionless. A thin lens of focal length $f_1$, numerical aperture $\eta$, and clear aperture $a$, collects scattered light from atoms in the lattice. A second magnifying thin lens of focal length $f_2 = \mu f_1$, where the magnification $\mu >> \ell$, forms the image of the target atom which is taken as the origin of the coordinate system.

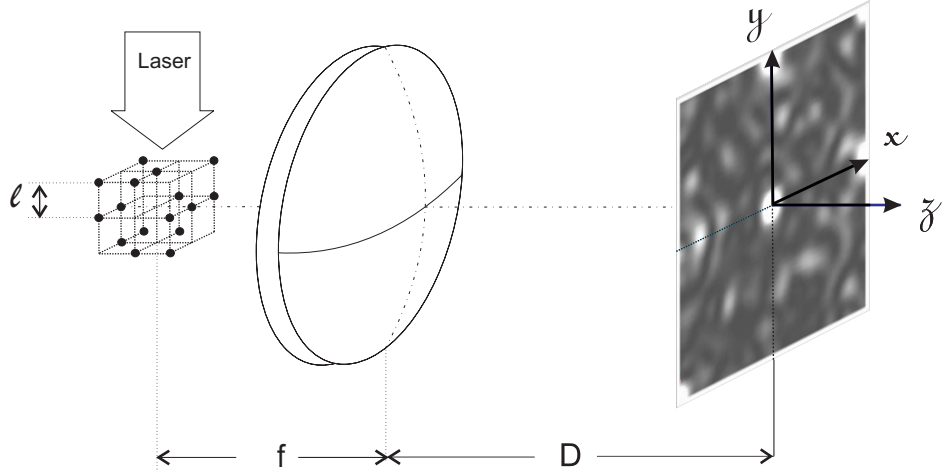

**Fig.1**: A schematic of the model for imaging the lattice.

We calculate the electric field near the origin due to all the atoms in the lattice, one plane at a time. Because the atoms radiate in a dipole pattern and the numerical aperture of the magnifying lens is small, all the imaged light has approximately the same polarization. We can therefore use the Fresnel-Kirchoff relation for the 3D distribution of the electric field as approximated by Debye [16] to calculate the amplitude in the $z = 0$ image plane due to an atom at the center of the $n^{th}$ lattice plane,

$$E(u_n, v_n) = -iA\frac{2\pi}{\lambda}\left(\frac{a}{d_n}\right)^2 \exp[i(d_n/a)^2 u_n] \int_0^1 J_0(v_n\xi) \exp\left[\frac{-iu_n\xi^2}{2}\right]\xi \, d\xi.$$

$J_0$ is a Bessel function, $A$ is a scaling constant, and $\xi$ is dimensionless and proportional to the lens radius. The subscript $n$ denotes the corresponding quantity for the image of the $n^{th}$ lattice plane, where $\mu_n = \mu/(1+(1+\mu)n\,\rho/f_1)$ is its magnification, $d_n$ is its axial distance from the magnifying lens and $z_n = d_n - d_0 = -n\,\rho\mu\mu_n$ is its z-coordinate. The field is expressed in terms of the scaled z-coordinate $u_n = 2\pi\chi^2\,\mu_n^{-2}(1+n\rho/f_1)^{-2}z_n$ and the scaled polar coordinate $v_n = 2\pi\rho\chi r/(1+n\rho/f_1)$ which both depend on the tangent of the aperture angle, $\chi = \tan(Sin^{-1}\eta) = a/f_1$, which differs from $\eta$ for large $\eta$. For the Debye integral to yield a good approximation to the light distribution near the focus, the values of the focal length, numerical and clear aperture need to be chosen so that $d_n >> a >> \ell$ and $a^2/d_n >> \ell$ for all values of $d_n$. Typically, $d_n$ is of the same order as $f_2$.

The amplitude $E_{00}^m(r) = E(u_m, v_m(r))$ for points in the $z = 0$ plane due to the atoms at (0, 0) sites in various lattice planes was generated for various $\eta$ and $\rho$ at a set of radii $r$. We use cylindrical symmetry to write the amplitude across the plane as $E_{00}^m(x,y) = E_{00}^m(r(x,y)) = E_{00}^m(\sqrt{x^2+y^2})$. If $\alpha_{pq}^m$ is the length of the path joining the lattice site at $(p, q)$ in the $m^{th}$ plane to its geometrical image, its image field in the $z = 0$ plane can be expressed as

$$E_{pq}^m(x,y) = \exp[2\pi i(\alpha_{pq}^m - \alpha_{00}^m)/\lambda]E_{00}^m(x+p, y+q).$$

The image of the atom at $(p, q)$ is at $(-p,-q)$. The expression for the intensity for a $(2N+1)^3$ lattice with the target site at the center can then be written as [17]

$$I_{net}(x,y) = \left|\sum_{m=-N}^{N}\sum_{p,q=-N}^{N}\beta_{pq}^m\,E_{pq}^m(x,y)\right|^2. \quad (1)$$

The occupancy $\beta_{pq}^m$ for each site was assigned the value 0 or 1 pseudo-randomly with a probability of 0.5. The intensity when $\beta_{00}^0 = 0$ ($\beta_{00}^0 = 1$) represents the image of the lattice when the target atom is absent (present). Fig. 2 shows contour plots of $I_{net}(x,y)$ for a typical random half-occupation of an $N=4$ lattice (9×9×9) imaged by an $\eta=0.51$ lens. We show plots for $\beta_{00}^0 = 1$ and 0 for two different $\rho$ values, 6.1 and 6.2, which are comparable to the one used in Ref. [6]. The average intensity in the two pairs of image planes differs greatly. Surprisingly, when $\rho$=6.2, the intensity at the image of the target decreases when the target atom is present. We will first expand on these numerical results, and then qualitatively explain them.

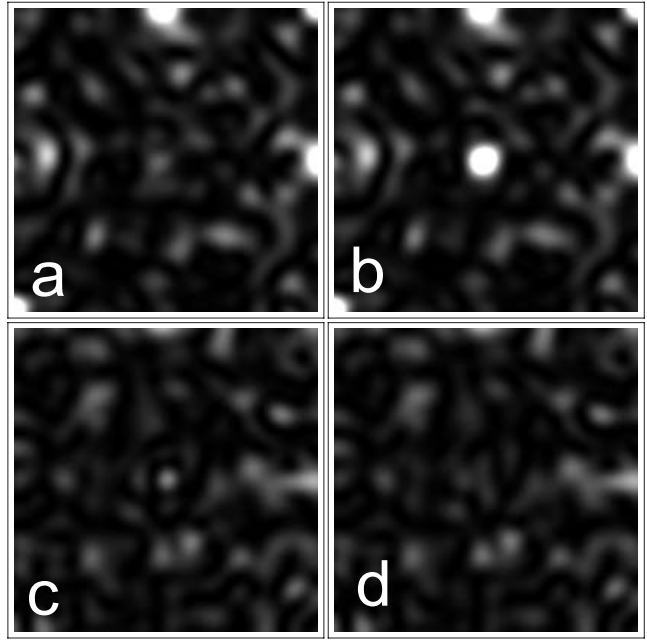

**Fig. 2**: Numerically calculated contour plots of 4 unit cells around the target site in the image plane. All images are for a half-filled $9^3$ lattice imaged with an η=0.514 lens. a) $\rho$=6.1, with no target atom. b) $\rho$=6.1, with a target atom. c) $\rho$=6.2 with no target atom. d) $\rho$=6.2 with a target atom. The same linear gray scale applies to all images. For $\rho$=6.1 the target intensity increases when a target atom is present. For $\rho$=6.2, the target intensity counterintuitively decreases when a target atom is present.

The ability to identify the presence of a target atom depends on how the intensity $I_{tgt} = I_{net}(0,0)$ at the corresponding image point (0, 0) changes with and without the target atom. To quantify this ability, we calculate $I_{tgt}$ in both cases for a set of 2000 random occupations of a cubic lattice with 9 sites per side. A typical set of results is plotted in Fig. 3. We define a threshold intensity $I_{th}$ such that over an ensemble of random occupations, $Prob(I_{tgt} > I_{th} | \beta_{00}^0 = 1) \equiv Prob(I_{tgt} < I_{th} | \beta_{00}^0 = 0)$, i.e., the probability of $I_{tgt}$ being above threshold when there is a target atom equals the probability of $I_{tgt}$ being below threshold with no target atom. The probability of incorrectly determining a site's occupation is $\varepsilon \equiv Prob(I_{tgt} > I_{th} | \beta_{00}^0 = 0) = Prob(I_{tgt} < I_{th} | \beta_{00}^0 = 1)$. $\varepsilon$ depends on $\eta$, $\rho$, and the size of the lattice, and can in some cases exceed 50%. We assume here that shot noise in the collected light is small compared to the fluctuations due to the 3D occupancy distribution. In practice, $\varepsilon$ can be improved by taking into account spatial intensity distributions and using occupancy information from out of focus planes [6].

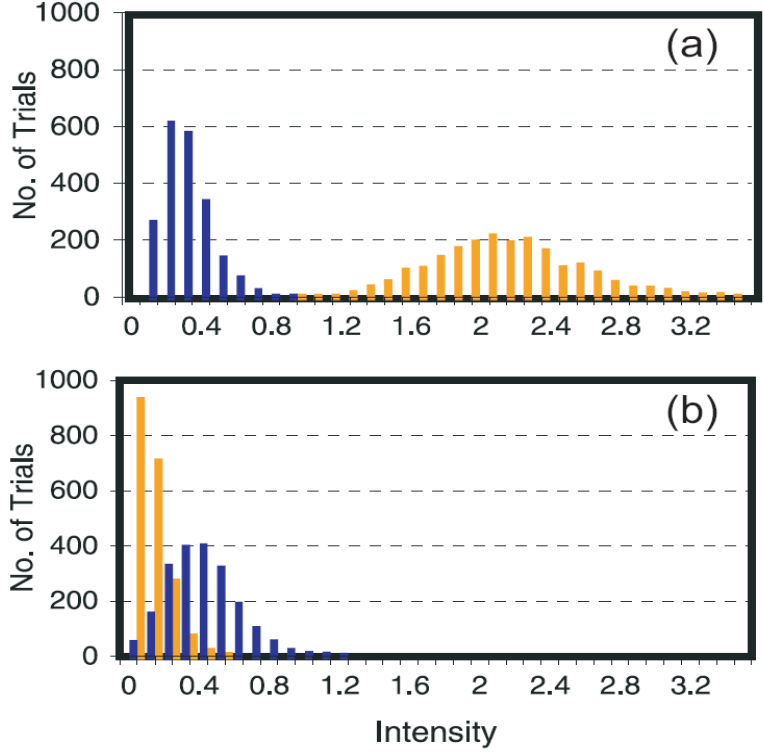

**Fig. 3**: Histogram of binned intensities at the center of the 2D image plane for 2000 randomly half-occupied $15^3$ lattices. a) $\rho = 6.1$. b) $\rho = 6.2$. The lighter (darker) bars correspond to a target atom being present (absent). For $\rho = 6.2$, the intensity tends to be higher in the absence of a target atom, which would make it impossible to resolve site occupancy.

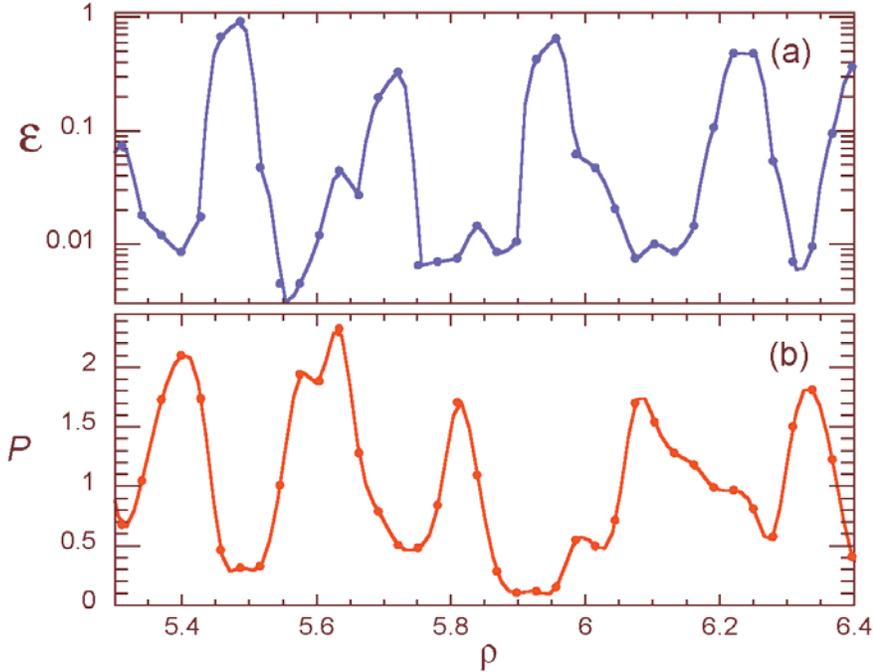

**Fig. 4**: a) The resolution error as a function of the rescaled lattice spacing, $\rho$, for $\eta=0.514$. The data is based on 2000 random distributions. b) The net power collected by the imaging lens as a function of $\rho$,. The data is for $\eta=0.514$ lens and a full $15^3$ lattice. The minimum (maximum) error points in (a) are associated with power peaks (valleys) in (b). The lines are to guide the eye.

Fig. 4a shows $\varepsilon$ as a function of lattice spacing for $\eta=0.514$ on a log scale. Site resolvability varies dramatically and aperiodically with $\rho$. The curve in Fig. 4a can be qualitatively explained by far-field interference among Bragg diffraction peaks, which is relevant because $f>>n\ell$. Fig. 4b shows, as a function of $\rho$, $P$, which is the calculated power collected by the imaging lens from a $15^3$ fully occupied lattice, normalized by the power that would be collected without interference. $P$ varies due to changes in the strength and angular location of Bragg diffraction peaks. Comparing Fig. 4a to 4b, we see that there is less error when $P$ is larger, and vice versa. The correlation can be understood as follows. The marginal effect of adding an atom in the target plane is to enhance interference effects. So when many Bragg diffraction peaks subtend the lens, adding a target atom disproportionately increases $P$. The entire change in $P$ is seen at the target site in the image plane. Conversely, when there are Bragg peaks that do not subtend the lens, or there is net destructive interference in all directions, adding an atom can decrease $P$. So even though all the light from the target atom that subtends the lens is focused near its image point, the total amount of light through that point can actually decrease.

One would expect that imaging details unrelated to Bragg diffraction, like $f_1$ or $f_2$, would only minimally affect $\varepsilon(\rho)$. Accordingly, we find that tripling $f_2/f_1$ has no impact on the location of the error features. For a given $\eta$, the minimum $\varepsilon$ increases by more than an order of magnitude as the lattice size increases from $9^3$ to $15^3$. Increasing $\eta$ tends to decrease the error, partly because of the sharper focus of light from the target atom in the image plane, but also because more Bragg peaks are collected. When there are many Bragg peaks just outside the lens, we find that, as $\eta$ is increased, $\varepsilon$ (at least for the central lattice site) decreases before $P$ increases. For instance, the $\rho$=5.9 point in Fig. 4, which is anomalous in that it has low $\varepsilon$ and low $P$, becomes a $P$ maximum when $\eta$ is made just 3% bigger.

We have also performed calculations with other occupancy fractions. The error rate improves away from half-filling, which is the highest entropy distribution. For nearly full lattices and a favorable $\rho$, site occupation can be determined with unmeasurably small error, even with $N$=7 (>3000 lattice sites).

In summary, we find that good resolution of well-localized single atoms in a 3D lattice requires maximum net Bragg diffraction into the imaging system. For a given size and orientation of lattice, the reliability with which site occupation can be determined depends sensitively on the ratio of the lattice spacing to the scattering wavelength, and relatively weakly on other parameters. These results can be applied to imaging neutral atoms in optical lattices as well as to imaging other ordered arrays of scatterers.

We acknowledge the support of the U.S. Army Research Office and DARPA.

**References**

1. G. K. Brennen, C. M. Caves, P. S. Jessen, and I. Deutsch, "Quantum logic gates in optical lattices", *Phys. Rev. Lett*. **82**,1060 (1999).

2. D. Jaksch, H. J. Briegel, J. I. Cirac, C. W. Gardiner, and P. Zoller, "Entanglement of atoms via cold controlled collisions", *Phys. Rev. Lett.* **82**, 1975 (1999).

3. D. Jaksch, J. I. Cirac, P. Zoller, S. L.Rolston , R. Cote, and M. D. Lukin, "Fast quantum gates for neutral atoms", *Phys. Rev. Lett.* **85**, 2208 (2000).

4. J. Vala, A. V. Thapliyal, S. Myrgren, U. Vazirani, D. S. Weiss, and K. B. Whaley, "Perfect pattern formation of neutral atoms in an addressable optical lattice", *Phys. Rev. A.* **71**, 032324 (2005).

5. D.S. Weiss, J. Vala, A.V.Thapliyal, S. Myrgren, U. Vazirani, and K.B. Whaley, "Another way to approach zero entropy for a finite system of atoms", *Phys. Rev. A* **70**, 040302(R) (2004).

6. K. D. Nelson, X. Li and D. S. Weiss, "Imaging single atoms in a three dimensional array", *Nature Physics*,.**3**, 556 (2007).

7. J.J. Bollinger, "Crystalline order in laser-cooled, non-neutral ion plasmas", *Phys*. *Plasmas* **7**, 7 (2000).

8. H. Walther, "From a single ion to a mesoscopic system—crystallization of ions in Paul traps", *Physica Scripta* **T59**, 360 (1995).

9. Y. Miroshnychenko, W. Alt, I. Dotsenko, L. Forster, M. Khudaverdyan, D. Meschede, D. Schrader, and A. Rauschenbeutel, "An atom-sorting machine", *Nature* **442**, 151 (2006).

10. S. Bergamini, B. Darquie, M. Jones, L. Jacubowiez, A. Browaeys, and P. Grangier, "Holographic generation of microtrap arrays for single atoms by use of a programmable phase modulator", *J. Opt. Soc. Am. B* **21**, 1889 (2004).

11. W.L. Bragg, "The diffraction of short electromagnetic waves by a crystal", *Proc. Cambr. Phil. Soc.*, **17**, 43 (1912).

12. G. Birkl, M. Gatzke, I. H.Deutsch, S.L. Rolston, and W. D. Phillips, "Bragg scattering from atoms in optical lattices", *Phys. Rev. Lett.*, **75**, 2823 (1995).

13. M. Weidemüller, A. Hemmerich, A. Gorlitz, T. Esslinger, and T.W. Hansch, "Bragg-diffraction in an atomic lattice bound by light", *Phys. Rev. Lett.*, **75**, 4583 (1995).

14. S.L. Winoto, M.T. DePue, N.E. Bramall, and D.S. Weiss, "Laser cooling at high density in deep far-detuned optical lattices", *Phys. Rev A*, **59**, R19 (1999).

15. [15]M. T. DePue, C. McCormick, S.L. Winoto, S. Oliver, and D.S. Weiss, "Unity occupation of sites in a 3D optical lattice", *Phys. Rev. Lett.* 82, 2262 (1999).

16. M. Born and E. Wolf, "Principles of Optics", (Cambridge, London, 2002), pp.484-487.

17. When scatterers are delocalized over a wavelength, interferences wash out and intensities from different atoms can be simply added. Because the intensity from an out-of-focus atom scales as $m^{-2}$ and the number of contributing atoms scales as $m^2$, each out of focus plane on average contributes approximately the same amount of light to the background. The signal to background ratio in that case is $\sim 2N^{-1}\pi\rho^2 \left(Sin^{-1}(\eta)\right)^4 (\chi)^{-2}$.